\documentclass[final]{scrartcl}
\usepackage{a4wide}
\usepackage{showkeys}
\usepackage{amsmath, amssymb, epsf}
\usepackage{graphics,graphicx}
\usepackage{algorithm}
\usepackage{algorithmic}
\usepackage{graphics,graphicx}

\begin{document}



\title{TrustMIX: Trustworthy MIX for Energy Saving in Sensor Networks}
\author{
 Olivier Powell$^1$\thanks{Research supported by the \emph{Swiss National Science Foundation} (SNF), ref. PBGE2 - 112864.}
\and Luminita Moraru$^2$\thanks{Swiss SER Contract No. 05.0030}
\and Jean-Marc Seigneur$^3$}
\date{\small
\noindent
$^1$Computer Engineering and Informatics Department, Patras University, Greece.
\\\noindent
$^2$Computer Science Department, University of Geneva, Switzerland.
\\\noindent
$^3$Information Systems Department, University of Geneva, Switzerland.
}
\maketitle

\begin{abstract}
\texttt{MIX} has recently been proposed as a new sensor scheme with better
energy management for data-gathering in Wireless Sensor Networks. 
However, it is not known how it performs when some
of the sensors carry out sinkhole attacks. In this paper, we propose
a variant of \texttt{MIX} with adjunct computational trust management to limit
the impact of such sinkhole attacks. We evaluate how \texttt{MIX} resists
sinkhole attacks with and without computational trust management.
The main result of this paper is to find that \texttt{MIX} is very
vulnerable to sinkhole attacks but that the adjunct trust management 
efficiently reduces the impact of such attacks while preserving the
main feature of \texttt{MIX}:
increased lifetime of the network.
\end{abstract}

\section{Introduction}
Wireless sensor networks have received a great deal of attention from the research community during the last decade. One of
the main challenges of sensor
network design is to save the limited available energy in order to ensure long term functioning.
A number of schemes have been proposed to minimize energy consumption at all layers, including the routing layer we are here interested in,see section \ref{Related work: energy saving}.
In a typical application, the network nodes have to propagate environmental information gathered by their sensors to base station.
Most of the time, sensors use short-range radio transmission to save energy since the required transmission power increases super-linearly with the distance between sender and receiver.
As a consequence, messages are propagated in a hop-by-hop fashion from the source to the base station and message routing becomes necessary. Many routing schemes have been proposed, typically, messages can follow a gradient established during the initialization phase of the network (nodes $n$ hops away from the base station send messages to nodes $n-1$ hops away and so forth).
In previous work, we have proposed one such gradient-based routing algorithm called \texttt{MIX} \cite{mix,JLPR06}. In \texttt{MIX}, we assume that sensors nodes can vary their transmission power, and thus their transmission range. This feature is used to send, occasionally, messages directly to the base station even if the base station is still quite far away. Typically, this can be used to avoid the last few hops around the base station where high traffic, congestion and high energy depletion can occur.
The fact of avoiding the last few hops by sending a message directly to the base station is referred as \emph{ejecting} the message. It has been shown in \cite{mix,JLPR06,PLR05,PLR06} that a few occasional but well employed message ejections can greatly increase the overall network lifetime.

The \texttt{MIX} algorithm works well when all sensors cooperate. However, in some real settings, the cooperation assumption may not be valid. For example, a few sensors may lie about their current energy level to avoid having to forward messages or worse they may not forward messages when asked to do so. In the latter case, these misbehaving sensors carry out an attack commonly called \textit{sinkhole attack} \cite{sinkhole}.
In this paper, we evaluate the impact of sinkhole attacks on
\texttt{MIX}. In addition, we propose and evaluate an adjunct trust management
mechanism to mitigate that impact.
The organisation of the paper is as follows. In section \ref{related work},
we survey the related work. In Section \ref{trustworthy MIX}, we
detail \texttt{MIX} and the model
for adjunct trust management. Section \ref{evaluation} describes our evaluation
testbed and presents our results. We draw conclusions and propose
future work directions in section \ref{conclusion} and \ref{future work}.

\section{Related Work}
\label{related work}
In this section, we first give an overview of what we mean by
computational trust management (section \ref{computational trust}). 
Then, we survey how it has been
used in the field of sensor network application (section 
\ref{trust in WSNs}) and in section \ref{Related work: energy saving} we
  give an overview of the main energy saving schemes for sensor networks.
\subsection{Computational Trust Management Overview}
\label{computational trust}
In the human world, trust exists between two interacting entities and
is very useful when there is uncertainty in the result of the
interaction. The requested entity uses the level of trust in the
requesting entity as a means to cope with uncertainty and to engage in an
action in spite of the risk of a harmful outcome. There are many
definitions of the human notion of trust in a wide range of domains, with
different approaches and methodologies: sociology, psychology,
economics, pedagogy. Romano's recent definition tries to encompass
previous work in all of these domains: \emph{Trust is a subjective assessment
of another's influence in terms of the extent of one's perceptions
about the quality and significance of another's impact over one's
outcomes in a given situation, such that one's expectation of,
openness to, and inclination toward such influence provide a sense of
control over the potential outcomes of the situation}
\cite{natureOfTrust}. 

\begin{figure}[hbt]
\begin{center}
\includegraphics[angle=0,width=.6\textwidth]{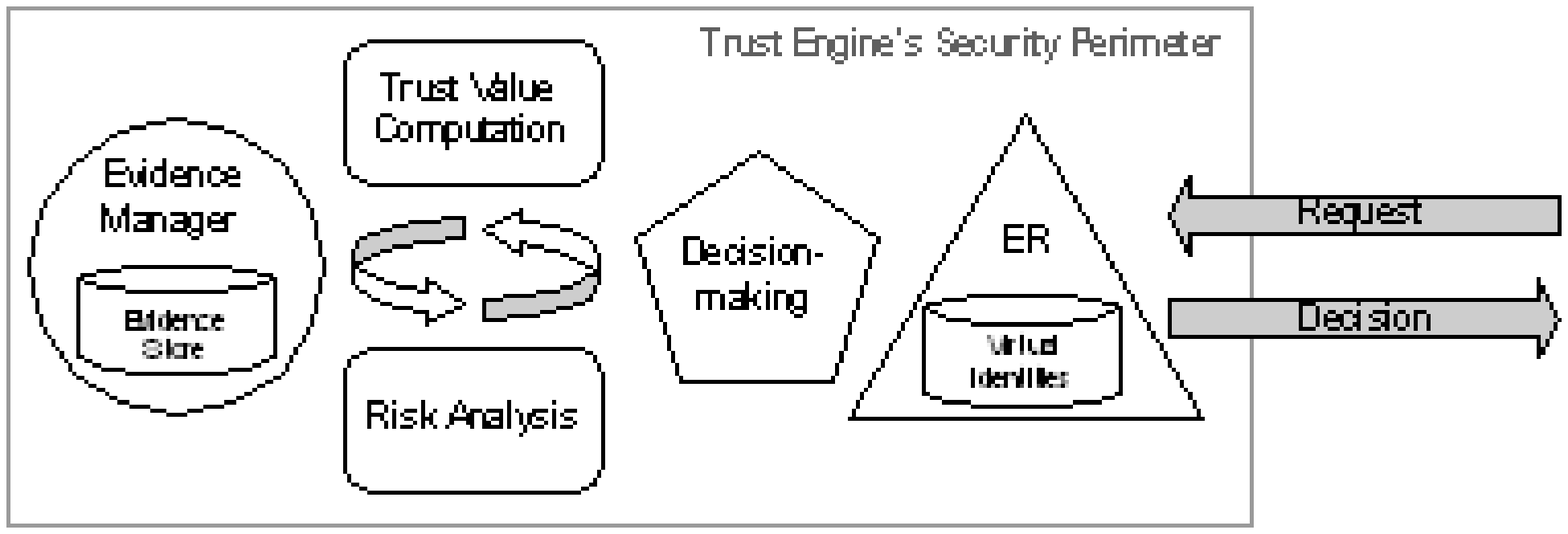}
\end{center}
\caption{High-level view of a trust engine}
\label{fig of trust engine}
\end{figure}

A computed trust value in an entity may be seen as the digital
representation of the trustworthiness or level of trust in the entity
under consideration: a non-enforceable estimate of the entity's future
behaviour in a given context based on evidence \cite{trustcomp}. A
computational model of trust based on social research was first
proposed by Marsh \cite{marsh94formalising}. Trust in a given
situation is called the trust context. Each trust context is assigned
an importance value in the range $\left[0,1\right]$ and a utility value in the range
$\left[-1,1\right]$. Any trust value is in the range $\left[-1,1\right)$. Risk is used as a
threshold for trusting decision making. Evidence encompasses the
outcome of
observations, recommendations and reputation. The propagation of
trust in peer-to-peer networks has been studied by Despotovic and
Aberer \cite{despotovic} who introduce a more efficient algorithm to
propagate trust and recommendations in terms of computational and
communication overhead. Such overhead is especially important in
sensor networks as any overhead implies more energy spending.  
A high level view of a trust engine is depicted in Figure \ref{fig of trust engine}. 
The decision-making component can be called whenever a trusting
decision has to be made. Most related work has focused on trust
decision-making when a requested entity has to decide what action
should be taken due to a request made by another entity, the
requesting entity. It is the reason that a specific module called
\emph{Entity Recognition} (ER) \cite{jmTrust} is represented to recognise any
entities and to deal with the requests from virtual identities.
The decision-making of the trust engine uses two sub-components:
a trust module that can dynamically assess the trustworthiness of the
requesting entity based on trust evidence of any type stored in
the evidence store;
a risk engine that can dynamically evaluate the risk involved in the
interaction, again based on the available evidence from the evidence
store.
A common decision-making policy is to choose (or suggest to the user)
the action that would maintain the appropriate cost/benefit. In our
sensor network application domain, we have to balance message ejection 
and forwarding rates,  based on how much energy has to be spent in
each case to successfully reach the base station. In the background,
the evidence manager component is in charge of gathering evidence
(e.g., recommendations, comparisons between expected outcomes of the
chosen actions and real outcomes). This evidence is used to update
risk and trust evidence. Thus, trust and risk follow a managed
life-cycle.

Although J\o sang's "subjective logic" does not use the notion of risk,
it can be considered as a trust engine that integrates the element of
ignorance and uncertainty, which cannot be reflected by mere
probabilities but is part of the human aspect of trust. In order to
represent imperfect knowledge, an opinion is considered to be a
triplet, whose elements are belief $b$, disbelief $d$ and uncertainty
$u$, such that \[ b+d+u=1 \textsl{ for }\left(b,d,u\right)\in\left[0,1\right]^3
\]
 		
The relation with trust evidence comes from the fact that an opinion
about a binary event can be based on statistical evidence. Information
on posterior probabilities of binary events is converted in the $b$, $d$
and $u$ elements to a value in the range $\left[0,1\right]$. The trust value $w$ in
the virtual identity $S$ of the virtual identity $T$ concerning the
trust context $p$ is:
\[ w^T_{p\left(S\right)}=\left\{b,d,u\right\}\]
The subjective logic provides more than ten operators to combine
opinions. For example, the recommendation ($\bigotimes$) operator corresponds to
using the recommending trustworthiness $RT$ to adjust a recommended
opinion. J\o sang's approach can be used in many applications since the
trust context is open. In this paper, we base our sensor trust values
on this kind of triple and statistical evidence count, c.f. section \ref{adjunct trust}.
\subsection{Trust and Security in Sensor Networks}
\label{trust in WSNs}
Wireless sensor
networks contain hundreds of entities used to collect data from the
environment where they are deployed. To save their their limited available energy, low
power and thus short range radio transmission is preferred.
As a consequence, each sensor
relies on its peers to forward collected data to a central entity, a
base station. Limited in energy,
sensors are also motivated to have a non-cooperative behaviour when it
comes to relaying packets from other sensors. They can save power by not
forwarding messages received from their neighbours. However, selfishness is
not the only misbehaviour that a sensor network has to cope with. 
An attacker can as well compromise sensors and prevent packets from
reaching their destination. Several types of attacks have been
identified. 
A sensor behaving like a sinkhole will drop any packet it
receives \cite{sinkhole}. In a wormhole attack \cite{hu02wormhole},
two colluding sensors create a tunnel between them. The first sensor is
situated in the proximity of the base station and relays the messages
received by the second one. The tunnel is a fast path and will
encourage the sensors to use it for routing. In a Sybil attack
\cite{sybil}, a sensor claims to have multiple identities. For a
routing protocol that uses several paths to the destination, a Sybil
attack can advertise one path as several ones. Additionally, it can be
correlated with sinkhole or wormhole attacks.

A first line of defense is the distribution of private keys to each
sensor. But sensors are low cost devices, without tamper proof
hardware, thus a captured sensor will permit access to its
cryptographic material. Key management schemes \cite{ebs,random} 
try to  increase network resilience to sensor capture
while maintaining the performance objectives and minimizing the
resulting cost of lower network connectivity due to sensors which do not
share similar secret keys. There is a trade-off between the energy
spent, the amount of memory used for protection and the security
level reached \cite{hwang04energymemorysecurity}. Static keying means
that sensors have been allocated keys off-line before deployment. 
The easiest way to secure a network is to give a
unique key at pre-deployment time but if only one sensor is
compromised the whole network is compromised and it seems viable to
extract the key from one sensor as they are cheap and not so well
protected (in non-military application scenarios). 
The second approach
is to have pair-wise keys for all sensors on each sensor, which is
impractical due to the memory constraints on the sensors. Dynamic
keying means that the keys can be (re)generated after deployment: it
creates more communication overhead but stronger resilience to sensor
capture. 
Radio transmission consumes most of the energy spent for
security mechanisms,  encryption only represents 3\% of the total energy
consumption \cite{hwang04energymemorysecurity}. 
Thus, minimizing security transmission is
important with regard to energy
saving. 

In \cite{reputationGaneriwal}, the authors argue that
previous approaches relying on
keying management and cryptographic
means are not suitable for sensors
due to their resource constraints or
the fact that it is easy to recover
their cryptographic material because
they are cheap and not fully
tamper-proof.
In this paper, we do not follow the cryptographic approach. Instead,
we focus on another way of detecting and preventing misbehavior by using \emph{computational trust management} as presented
section \ref{computational trust}. 
Several mechanisms have been already proposed for mobile
ad-hoc networks and wireless sensor networks. They are concerned with
making decisions on whether to cooperate or not with their peers based on
their previous behaviour. The information used to build the reputation
value of neighbours is collected mainly by direct interaction and
observation. Although it is accurate, it requires some time before
enough evidence has been collected. In our scenario, where sensors are
static, there is more time to build trust among neighbour
sensors since they do not move. 
If recommendations are used, the reputation of the sensors that
provide the recommendations has to be taken into account. In this
latter case, it may also  generate vulnerabilities to false report
attacks.

Existing trust models have different approaches to such reputation
building and decision making problems. CORE \cite{core} builds the
reputation of a sensor as a value that is increased on positive
interactions and decreased otherwise. It also takes into account
positive ratings from the neighbours. If the aggregated value of the
reputation is positive, the sensor cooperates, otherwise it refuses
cooperation. CONFIDANT \cite{confidant} considers only negative
ratings from the neighbours. In order to compute a reputation value,
different weights are assigned to personal observation and reported
reputation. RFSN \cite{rfsn} uses only positive ratings and models the
reputation value as a probabilistic distribution, by means of a
beta distribution model. A sensor will cooperate with the neighbours
that have a reputation value higher than a threshold. In
\cite{sinkhole}, computational trust based
on direct observations to mitigate both sinkhole and wormhole
attacks is used. However, this work only covers the Mobile Ad-hoc NETwork
(MANET) Dynamic Source Routing (DSR) protocol
\cite{johnson94routing}. They cover two trust contexts: TPP, Packet
Precision for wormhole; and TPA, Packet Acknowledgment for
sinkholes. They combine the two trust contexts. If a sensor is
suspected to be a wormhole, the combined trust value $T$ is $0$. Otherwise
TPP is equal to $1$. TPA is a counter that is incremented each time a
sensor is used to forward a packet and an acknowledgement has been
received before a timeout; it is decreased otherwise. The inverse of
the combined trust value simply replaces the default cost of $1$ in the
LINK CACHE of the standard DSR protocol. If it is a wormhole the cost is set to infinity.

In this paper, we are interested in evaluating how the \texttt{MIX} 
data-gathering protocol for sensor networks, which is different from the protocols
covered by previous work on sensor trust management, is impacted by sinkhole
attacks with an emphasis on energy. In section \ref{adjunct trust} we
present the computational trust management model we have chosen. One
of the main requirement is to maintain the \texttt{MIX} energy saving
property while guaranteeing efficient delivery of messages under
sinkhole attacks using trust management.
\subsection{Energy Saving Schemes for Sensor Networks}
\label{Related work: energy saving}
In Wireless Sensor Networks (WSN), one of the main constraints
is to minimize energy consumption in order to maximize the
lifespan of the network. Indeed, sensor nodes are usually battery
powered \cite{RAS+00,WLLP01,SL05}. 
In order to increase the lifetime of sensor networks, various energy
saving schemes have been proposed. For example,
multi-hop transmission techniques \cite{IGE00,CNS02}, clustering
techniques \cite{HCB00}, alternating power saving modes \cite{STGS02}, varying
transmission levels with route selection \cite{CT00}, energy replenishment
\cite{LSS05}, multi-path routing \cite{HGWC02}, probabilistic forwarding
techniques \cite{BCN05}, mobile sinks \cite{L06,LH05,LH06}, varying
battery levels \cite{SD05} or varying transmission ranges \cite{OS06} are
among possible approaches. For survey papers the reader may want to look at \cite{B05,BN04,AK05,AY05}.
Another possibility for network lifetime maximization is energy
balancing, which is the approach we are interested in in this paper.
In an energy balanced network, all the motes deplete their energy at
the same rate. 
To our knowledge,
energy balancing was first proposed in \cite{SP03} to implement a sorting
algorithm in WSN's, \cite{YP03} to solve a task allocation problem and in
\cite{GLW03} to implement a data-gathering algorithm for WSN's. 

In this paper, we are untested in sensor networks accomplishing a data-gathering task,
arguably the most common task for sensor networks. 
A data-gathering WSN is deployed over a region to be monitored and when
a mote detects an event, it needs to report to the sink. To do so, it
is usual \cite{BN04,AK05,AY05} to send messages to the sink in a hop-by-hop
fashion, from node to node towards the sink. While this minimizes the
\emph{total} energy spent by the network, since the cost
of sending a message from a mote to another grows like a power
(greater or equal to 2) of the distance between motes
\cite{AK05,AY05}, this type of data-propagation leads to premature energy depletion of nodes close to the
sink \cite{GLW03,LXG05,ENR04,ENR06,LNR05,PLR06,JLPR06,SD05,OS06}. 
Indeed, nodes close to the sink are
bottleneck nodes through which every message propagated through the
network has to pass, when using a hop-by-hop routing algorithm. 
This heavy load of traffic on motes close to the sink brings them to
deplete their energy rapidly.
Unfortunately, when too many of those nodes run out of energy, the sink
becomes disconnected from the network, thus putting the network down
while there may be plenty of energy remaining in motes away from the sink.
It thus seems that energy balancing is a particularly promising way
of maximizing the lifespan of networks accomplishing a data-gathering task.
We next present briefly the state of the art for this approach.

In
\cite{GLW03}, and then in \cite{LXG05}, the same authors propose an
energy balancing
data-propagation algorithm based on a division of time in two
different phases: during the first phase, sensors send data directly
to the sink while, in a second phase, sensors send data in a
hop-by-hop fashion. The duration ratio between the two phases is
critical to ensure energy balance, and the adequate value is found
through simulation.  In \cite{ENR04} and then in \cite{ENR06}, the authors
propose to use a probabilistic algorithm to ensure energy balance. The
network is divided into slices of sensors situated at the same hop distance
to the sink in the Unit Disc Graph (UDG) associated to the WSN. The
authors show that energy balance is achieved if a recurrence relation
is satisfied between the probabilities, for each slice, that sensors
eject messages directly to the sink. 
In \cite{LNR05} a more general
case is considered,
where the distribution of events is not known a priori but
dynamically inferred by the base station by observation of the
network. The optimal ejection probabilities for each slice are computed
and broadcast in the network periodically. In \cite{PLR06}, an algorithm is
proposed to compute offline the parameters maximizing the lifespan of
the network which is equivalent to the one from \cite{ENR04,ENR06} when an
energy balanced solution exists. However, they extend the work from
\cite{ENR04,ENR06} by also considering the case where the distribution of
events and the topology of the network lead to a situation where an
energy balanced solution does not exist, a case which had not been
considered in previous work, and prove analytically that their
algorithm produces an energy balanced and optimal solution when such a solution
exists, and that in other cases the algorithm
still outputs an optimal solution, although not energy balanced.
Previous algorithms were
useless when no energy balance solution exists. We would
like to emphasize that all these approaches use a centralised
algorithm to compute the optimal ratio of direct ejection to the sink
and hop-by-hop sliding of messages. 
In \cite{JLPR06}, the \texttt{MIX} algorithm was proposed. Similarly to previous work, \cite{JLPR06} considers that sensors may vary their transmission range. Short range is used to send messages to 1-hop away neighbours and occasional long-range transmission is used to send (or ``eject'') messages directly to the sink, typically to avoid the last few bottle-neck hops around the sink forming a ``hotspot'' in the network. \cite{JLPR06} also considers the possibility of using ``medium range'' transmissions. As an example, a message could be sent half-way to the sink. I.e., if the message is being sent by a sensor node located $n$ hops away from the sink, a medium range transmission could send the message to a sensor node $n/2$ hops away from the sink. Medium range transmissions could be interesting, although they imply a number of technical challenges (e.g. how can a message be routed 5 hops away if the node only knows its 1 hop away neighbourhood?), however, \cite{JLPR06} proves the surprising result that \emph{medium range transmissions are useless} in the sense that they cannot increase the lifetime of a network using a combination of short range transmission and long range ejections directly to the sink.
We describe in
detail the \texttt{MIX} algorithm in section \ref{description of MIX}.
\section{Trustworthy MIX Model}
\label{trustworthy MIX}
The \texttt{MIX} algorithm, proposed in \cite{JLPR06}, is a distributed algorithm
meant to maximize the lifespan of a wireless sensor network
accomplishing a data-gathering task. In this scenario, sensor nodes
are deployed over a region where some phenomenon is being
monitored. When an event is detected by one of the motes, it
data has the choice between sending a message directly to the sink (in
this  case
we say the message is \emph{ejected} to the sink) or sending it to one of
its neighbour nodes (in which case we say the message is \emph{slid}
along the network). The
mote which receives the message will be, in turn, faced with the same
choice: eject or slide the message. The 
ejection feature of the \texttt{MIX} algorithm lowers the traffic load
on nodes close to the sink, thus preventing the unwanted early energy
depletion of sensors close to the sink described in section \ref{Related work: energy saving}.
\subsection{Sensor Network Model}
\label{network model}
\paragraph{Communication graph}
We take the common approach of considering a WSN composed of $N$, a
set of sensors represented by
points in the Euclidean plane $\mathbb{R}^2$. One of them is a
distinguished node $S$ representing the sink. 
We use the unit disc graph (UDG), 
probably one of the most common communication graph models for sensor networks.
The UDG is built by adding an
edge between any pairs of sensors which are at distance at most
$1$ from one another. 
An edge represents a wireless link between two sensors (or the sink).
It may be worth pointing out that the unit disc graph is an idealized communication graph model and that recently it has been emphasized that relying too much on the model can lead to erroneous results. In this work, we rely on the communication graph for simulation purposes only and there seem to be no reason to believe that the adjunct trust scheme we propose in this paper behaves differently in other real or modeled communication graph. We are thus confident that the UDG assumption is harmless in the context of this paper.
\paragraph{Energy consumption model}
As explained, sensors can either send data to a neighbour node in the
UDG communication graph, or alternatively send data directly to
the sink, which is the message ejection feature of \texttt{MIX}.
We arbitrarily consider that a sensor $n$ sending a message to one of its
neighbours requires the sending sensor to spend $1$ energy unit,
whereas sending a message directly to the sink makes it spend $hop(n)^2$
energy units, where $hop$ is the length of the shortest path  from the sending node to
the sink in the UDG (for simplicity we make the assumption that the UDG is
connected).
In other words, we consider that the attenuation factor for increasing
transmission range is $2$, and we make the pessimistic approximation
that a sensor node $n$ is at Euclidean distance $hop(n)$ from the sink. As was
explained in \cite{ENR04,ENR06}, the arbitrary choice of $2$ for the
attenuation factor does not modify the global behavior of ejecting
algorithms (although it changes numerical values in simulation).
\paragraph{Attacker model}
In order to simulate the sinkhole attacks,
we consider that a subset $A$ of the sensor motes $N$ is composed of attacker motes. 
When receiving a message, those motes simply discard it,
and in order to capture as many messages as
possible, the attacker motes lie about their \emph{remaining energy} as well
as their \emph{hop distance}, pretending they have plenty of energy left (in
the simulations, we set their energy level to infinity) and that they
are closer to the sink than any of their neighbours. Thus, the
attacker motes apply a strategy which is directly aimed at fooling the
\texttt{MIX} algorithm 
since \texttt{MIX} considers motes to be good
relays for data propagation precisely when they have plenty of energy
left, and when they are close to the sink in terms of the $hop$
distance, c.f. section \ref{description of MIX} for a detailed description of \texttt{MIX}.

\subsection{The Basic MIX Energy Saving Scheme} 
\label{description of MIX}

Cite \cite{PN07,PN07b}

The \texttt{MIX} algorithm \cite{mix,JLPR06} is a
gradient based routing algorithm: when a
node needs to send a message it looks for its "lowest" neighbour and
sends it the message. If a node is located at a local minimum, it ejects
the message directly to the sink, thus the ejection feature. 
In \texttt{MIX}, the potential function is such that a node $m$ is
considered \emph{lower} than a
node $n$ if \[hop(m) < hop(n)\] or if \[hop(m) = hop(n)\textsl{ and }energy(m) <
energy(n)\] where $energy(m)$ is the energy consumed by node $m$ so
far and $hop(m)$ is the hop distance from the mote to the sink.
We give a more precise description of the \texttt{MIX} algorithm in the
pseudocode of figure \ref{MIX algorithm}.
\begin{algorithm}
\caption{\texttt{MIX} Routing algorithm for sensor $n$}{ }
\label{MIX algorithm}
\begin{algorithmic}[1]
\STATE \COMMENT{When $n$ gets a message, it will send it to $nextMote$.}
\STATE n = \texttt{Sensor running the algorithm}
\STATE nextMote = \texttt{``not defined''}
\FORALL{node $m$ among neighbours of $n$}
  \IF{$hop( m ) \geq hop( n )$}
    \STATE next
  \ELSIF{$energy( m ) \geq energy( n ) + hop^2( n )$}
    \STATE next
  \ELSIF{ $nextMote$ is \texttt{"not defined"}}
    \STATE nextMote = m 
  \ELSIF{$energy( m ) < energy( nextMote )$}
    \STATE nextMote = m 
  \ELSIF{ $energy( m ) = energy( nextMote)$}
    \STATE\COMMENT{Flip a coin to sort ties.} \\
    \STATE $nextMote= m$ with probability $0.5$ 
  \ENDIF
\ENDFOR
\IF{$nextMote = \texttt{``not defined''}$}
  \STATE\COMMENT{If $nextMote$ is not defined, we need to eject the message to the sink.}
  \STATE $nextMote = sink$
\ENDIF
\STATE\textbf{return}{$(nextMote)$}
\end{algorithmic}
\end{algorithm}

It may be noticed that in order to run on a mote, i.e. to make
\texttt{MIX} fully distributed, it is required that
each node has access to the remaining energy and hop distance of each
of its neighbours.
Each node is assumed to be aware of its own remaining energy (via its
embedded electronics), and each node is assumed to know its own
hop distance to the sink. This could typically follow from the
initialisation phase of the network,
during which a single flooding occurs from the sink. 
This implies that every
node sends one message (assuming no collisions occur), although
optimisation is possible, c.f. \cite{LMPR06} for a proposition of our own.
Knowing the hop distance of neighbours is easy,
since it is a constant value in a static network. However, even in  a
static network, the energy values change. Therefore, keeping aware of
the remaining energy of neighbour nodes, or at least
an estimation of this value, will require some extra care.
As explained in \cite{JLPR06}, this could for example be implemented
in a real WSN using standard piggy backing techniques, for example by including in
the header of messages the remaining energy and ID of the
sender. 
\subsection{Adjunct Trust Scheme}
\label{adjunct trust}
The trust scheme we propose is based on a trust function locally
computed at the sensor level and using belief, disbelief and
uncertainty values $\left(b,d,u\right)\in\left[0,1\right]^3$ for each pair
of nodes, as explained in section \ref{computational trust}. 
We make the
assumption that at each moment in time, for each node $m$ and for each
one of its neighbours $n$,  $m$  knows the number $s(m,n)$ of messages it has
sent to $n$, as well as $r(m,n)$ and $c(m,n)$ the number of messages $m$ has
sent to $n$ which have been received by the sink and captured by one
of the attacker motes respectively. In order to know the value of
$s(m,n)$, node $n$ simply needs to count the messages it sends to
$m$. However, keeping track of $r(m,n)$ and $c(m,n)$ may be
more difficult in  a real WSN (in our simulations, those values are
assumed to be available). Basically, what is required is some sort of transport layer protocol that permits to control when and if a message reaches destination.
One way of implementing this in a real network would be to make the
sink send acknowledgements (ACKs) when it receives a message. By
broadcasting from time to time a list of (hash keys) of received
messages, for example using a time division multiple
access scheme (TDMA), the sink could let all sensors become aware of the safe reception
of messages. This means however that
sensors need to store hash tables of messages awaiting an ACK and need to
listen for the ACKs from the sink, so there is an overhead in energy
consumption and memory usage. 
We would like to stress that
the long range broadcasting of ACKs by the sink is not a
problem in scenarios where the sink has plenty of available energy, 
for example when the sink is plugged on the
electrical network. Also, to prevent the attackers from sending
tainted ACKs, a static keying cryptographic scheme could be used,
c.f. section \ref{trust in WSNs}. 
Finally, in order to be aware of the number of captured messages
$c(m,n)$,
the node $m$ may simply set a time stamp for each message awaiting an
ACK from the sink. If the time stamp is reached, the message is
considered captured. With this approach, the belief $b$, the disbelief
$d$ and the uncertainty $u$ that node $m$ holds towards $n$,
c.f. section \ref{computational trust},  are defined as
$b=\frac{r(m,n)}{s(m,n)}$, $d=\frac{c(m,n)}{s(m,n)}$ and $u=1 - \frac{r(m,n) + c(m,n)}{s(m,n)}$.
Finally, we define the
trust that node $m$ has in $n$ to be
\[
  trust(m,n) 
  = \frac{ s(m,n) + r(m,n) - c(m,n) + 1 }{ s(m,n) + r(m,n) + c(m,n) + 1 }
\]
\[
  \stackrel{s(m,n)>0}{=} \frac{ 2b + u + \frac{ 1 }{ s(m,n) } }{ 2b + 2d + u + \frac{ 1 }{ s(m,n) } }
  \stackrel{s(m,n)\rightarrow\infty}{\longrightarrow} \frac{ 2b + u }{ 2b + 2d + u }
\]
The ``$+1$'' term in the definition above is just used to avoid
confidence dropping to $0$ in the case where a single message has
been sent and captured (i.e. when $s=c=1$ and $r=0$), and it leads to a vanishing term when
$s(m,n)$ tends to infinity.
The \texttt{trustMIX} algorithm we propose is a generalisation of the 
\texttt{MIX} algorithm using the above trust function. 
The idea in \texttt{trustMIX} is that when a node $n$ considers
neighbours to which it could forward
messages, it will refuse to take into consideration those it distrusts.
More precisely, when a sensor node loops over neighbours to which it
would possibly forward a message (i.e. while in the \texttt{for} loop of
algorithm \ref{MIX algorithm}), the sending node $n$ simply disqualifies
any neighbour $m$ with probability $1-trust(n,m)$. \texttt{trustMIX}
is thus a randomized algorithm. 
Our explanations are made precise in Algorithm \ref{trustMIX algorithm}.

\begin{algorithm}
\caption{Routing algorithm for sensor $n$}
\label{trustMIX algorithm}
  \begin{algorithmic}
  \STATE n = \texttt{Sensor running the algorithm}
  \STATE nextMote = \texttt{``not defined''}
  \FORALL{ node $m$ among neighbours of $n$ }
    \STATE $x = random( 0 , 1)$
   \IF{ $x > trust( n, m )$ }
      \STATE next
    \ELSE
      \STATE\COMMENT{Carry on with the main loop of the \texttt{MIX} algorithm}
      \IF{$hop( m ) \geq hop( n )$}
        \STATE next
      \ELSIF{$energy( m ) \geq energy( n ) + hop^2( n )$}
        \STATE next
      \ELSIF{ $nextMote$ is \texttt{"not defined"}}
        \STATE nextMote = m 
      \ELSIF{$energy( m ) < energy( nextMote )$}
        \STATE nextMote = m 
      \ELSIF{ $energy( m ) = energy( nextMote)$}
        \STATE $nextMote= m$ with probability $0.5$ 
      \ENDIF     
    \ENDIF
  \ENDFOR
  \IF{$nextMote = \texttt{``not defined''}$}
    \STATE nextMote = sink
  \ENDIF
  \STATE \textbf{return}{(nextMote)}
  \end{algorithmic}
\end{algorithm}

Please notice that in the case
where the network is trustworthy, the algorithm is identical to the
original \texttt{MIX} algorithm since the trust between pairs of
sensors is $1$. 
As we shall see in the evaluation
section \ref{evaluation}, 
simulations show that \texttt{trustMIX} manages to side-step sinkholes and to deliver a
significant fraction of messages to the sink, while preserving the
main feature of \texttt{MIX}: increased lifespan of the network.

\section{Evaluation}	
\label{evaluation}
\label{simulations}
We analyse our algorithm through simulations.
To conduct simulations, we consider a circle of radius $r$ and
randomly and uniformly scatter $n$
sensor nodes in the circle. A sink $S$ is placed at the center of the
circle. Optionally, $a$ attacker motes performing a sinkhole attack may
be scattered (randomly and uniformly) in the network. We divide time in rounds and each round
is divided in two phases: During the \emph{event detection phase}, we
randomly pick a non-attacker mote and make it detect an event by
incrementing the size of his message stack by one. During the \emph{message
propagation phase}, each sensor with a non-empty message stack sends
exactly one message. The sending mote runs algorithm \ref{trustMIX
  algorithm} to decide to which neighbour the
message is going to be slid, or if it should eject the message directly
to the sink. For simulation purposes, we use the following parameters:
$r$ is the radius of the circle in which nodes are initially
deployed, $d$ is a density factor and $p$ is the fraction of
attacker motes added. Given $r$, $d$ and $p$, we set $n$ to $n = \pi\cdot r^2\cdot d$
and $a = p \cdot n$. In order to evaluate the performances of our proposed
algorithm, we run in parallel and on the \emph{same network topology} and \emph{same
generated events} three scenarios. In the first scenario (1), no attackers
are placed in the network and the motes run the \texttt{MIX} algorithm
\ref{MIX algorithm}.
In the second scenario (2), the attacker motes are deployed and the \texttt{MIX}
algorithm is used again, without adjunct trust management (i.e. the
attack will be fully effective). Finally, the third
scenario (3) uses the adjunct trust scheme described in section
\ref{adjunct trust} and the motes run the \texttt{trustMIX} algorithm
\ref{trustMIX algorithm} as a counter-measure to the sinkhole attack.
\subsection{Basic MIX attack results}
\label{basic mix result}
\begin{figure}[hbt]
\begin{center}
\includegraphics[angle=0,width=.3\textwidth]{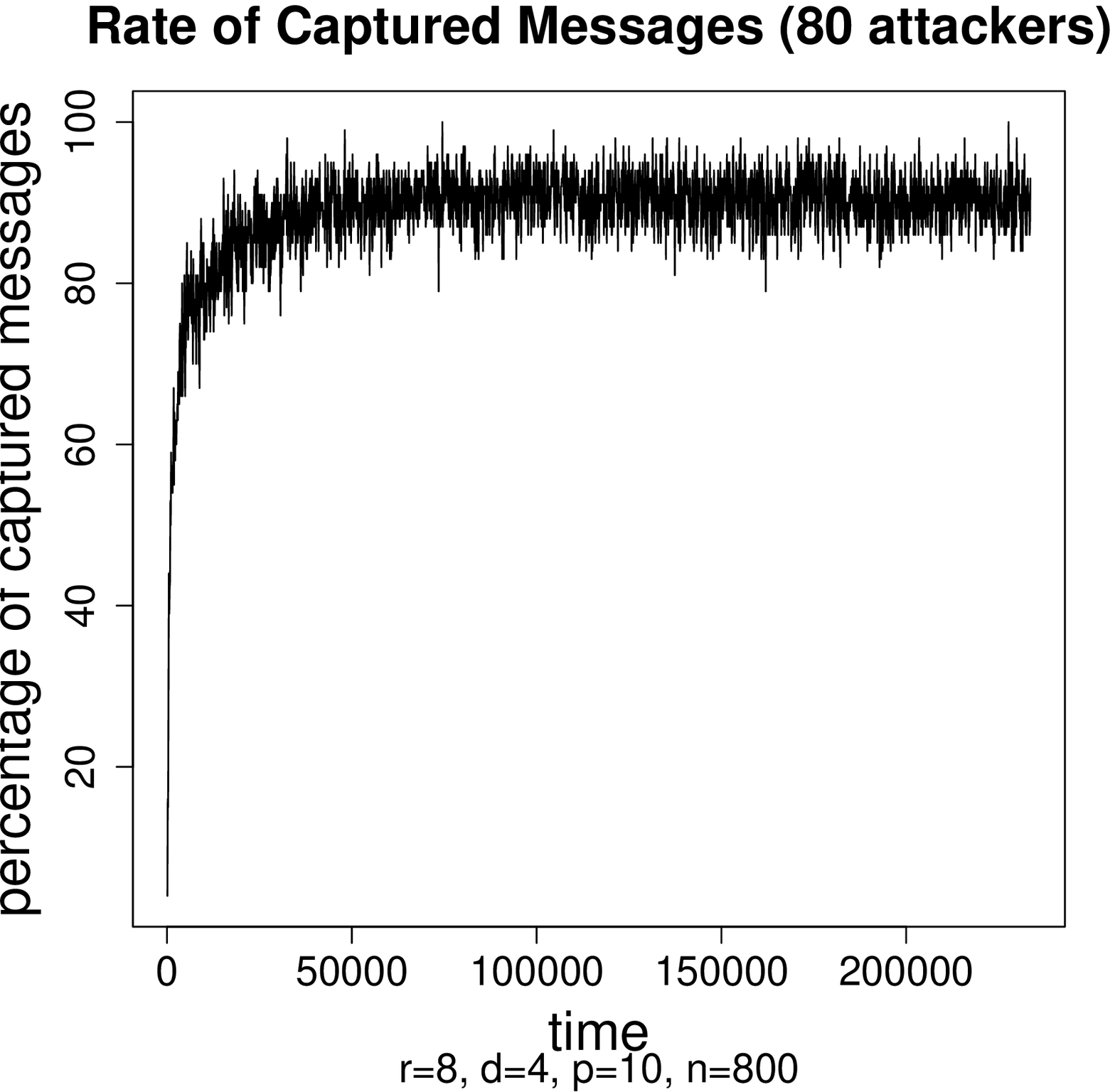}
\includegraphics[angle=0,width=.3\textwidth]{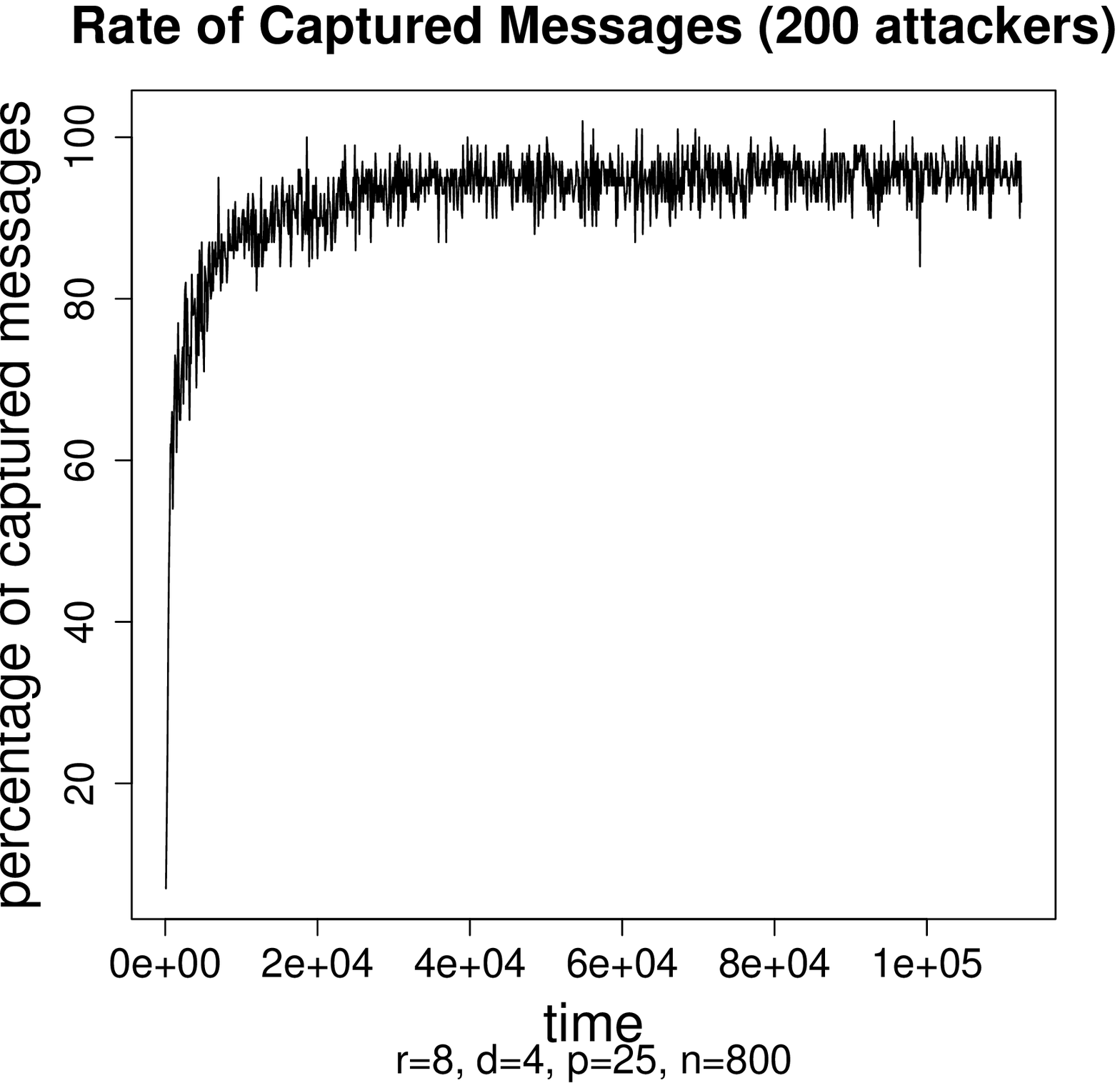}
\includegraphics[angle=0,width=.3\textwidth]{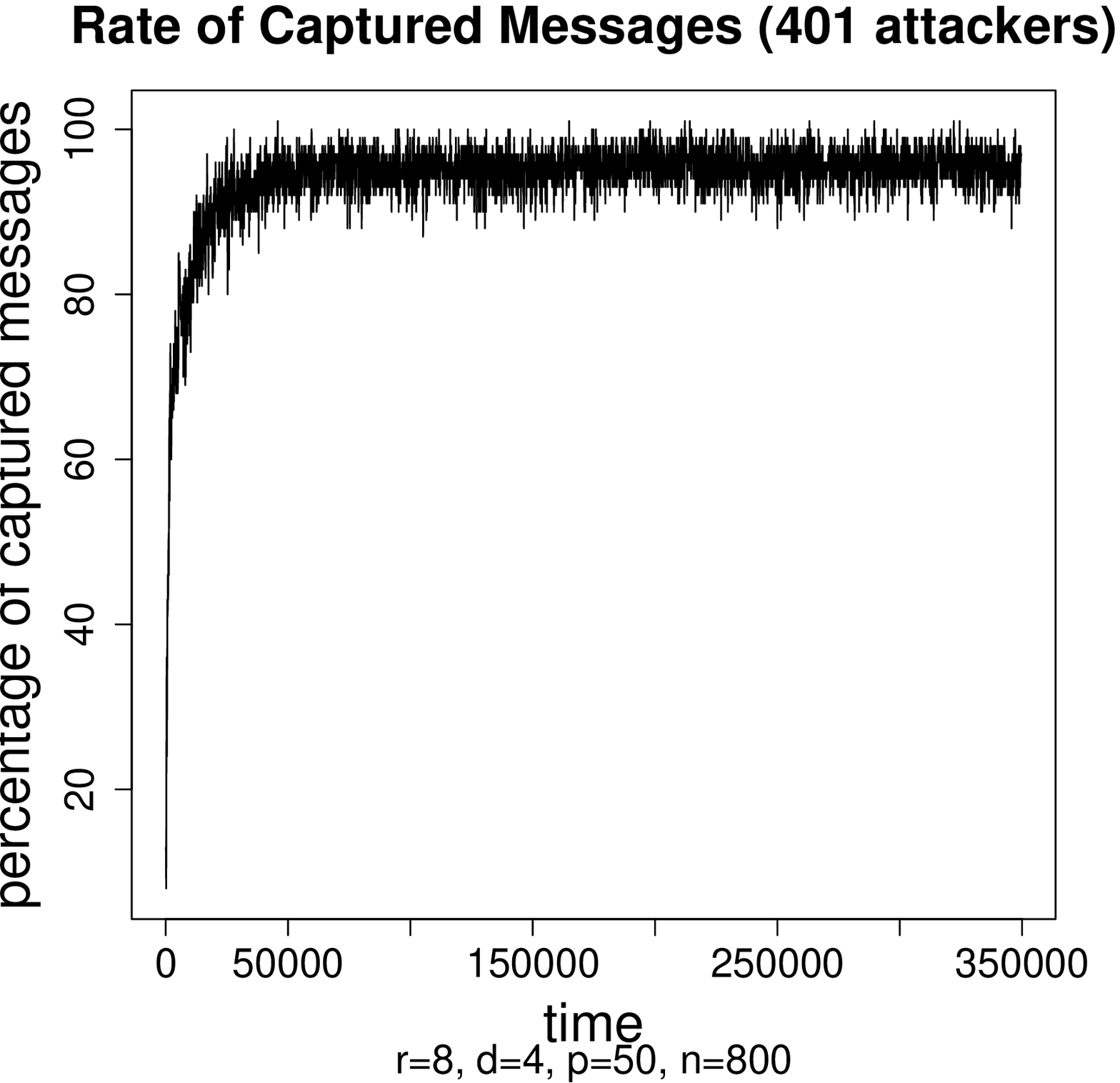}
\end{center}
\caption{Impact of sinkhole attacks on \texttt{MIX}}
\label{capture rates}
\end{figure}

We observe that the \texttt{MIX} protocol is vulnerable to sinkhole
attacks (scenario 2). The three plots of figure \ref{capture rates} show the result of our simulations when
$r$ and $d$ are set to $8$ and $4$ respectively, and when $p$ takes values in
$10$, $25$ and $50$. We observe that even with a
relatively small number of attacker motes (left-hand side plot of figure \ref{capture rates}),
attacker motes advertising themselves as having plenty of energy and
being close to the sink
rapidly become attractive to the \texttt{MIX} routing protocol, and soon manage to
hijack an large proportion ($80-90\%$) of the total
traffic.
Figure \ref{capture rates} shows the instantaneous
arrival rates at the sink.
In fact, the curve is smoothened
by taking the average over $100$ time steps. More precisely, since one message is generated in the network
at each time step, the smoothened instantaneous percentage of arrived
messages at time $t$ is defined as $f(t)$, with $f(t)=\sum_{i=0}^{99}
S_{rcv}(t-i)$, for each value of $p=10$, $p=25$ and
$p=50$, and where $S_{rcv}(t)$ counts messages received by the sink at
time $t$.
\subsection{MIX with Trust Management Attack Results}
The trust engine we use builds confidence levels between pairs of
nodes. Trustworthy motes are more likely to be selected for message
forwarding, whereas attacker motes will be avoided. The aim of the
\texttt{trustMIX} algorithm is to detect and avoid attacker nodes and to run the normal \texttt{MIX}
algorithm from \cite{JLPR06} on the remaining trustworthy motes. 
Success for \texttt{trustMIX} would mean that
messages get delivered to the sink while preserving the main feature of the
\texttt{MIX} algorithm: increased lifetime for the network. 
We shall show that this is indeed the case.
\subsubsection{Evaluating the Attenuation of the Sinkhole Attack}
\begin{figure}[hbt]
\begin{center}
\includegraphics[angle=0,width=.45\textwidth]{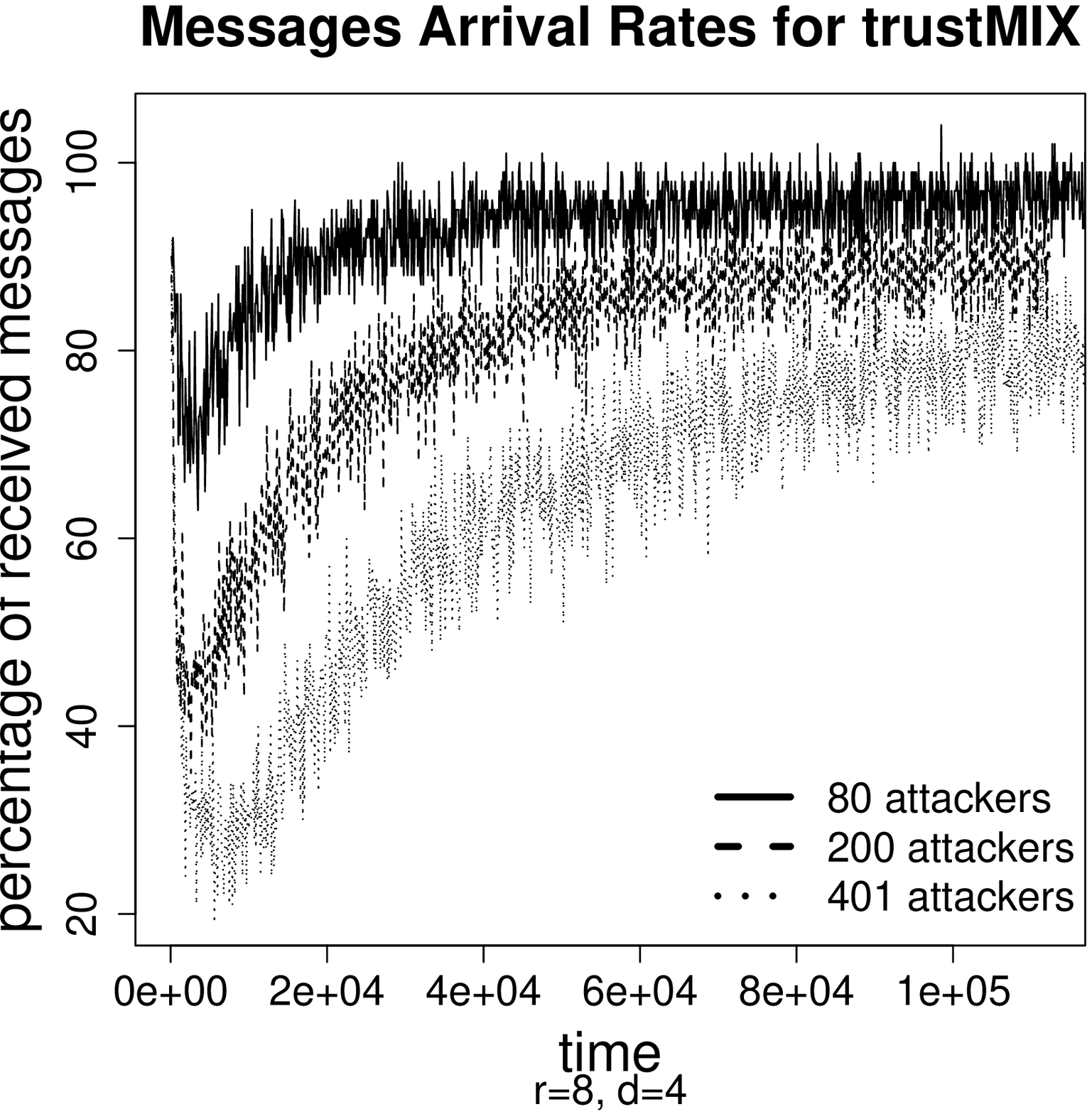}
\includegraphics[angle=0,width=.45\textwidth]{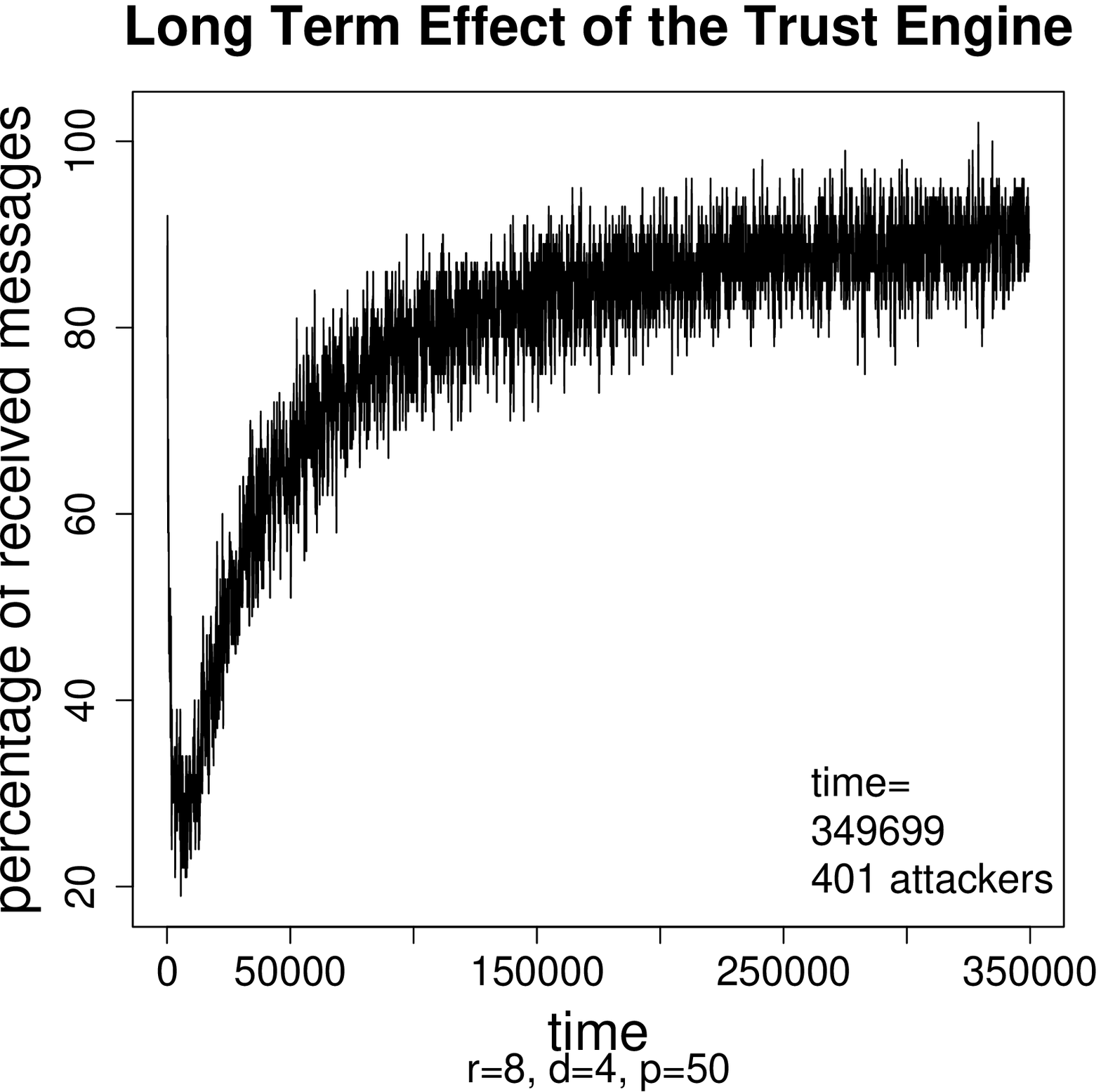}
\end{center}
\caption{Impact of \texttt{trustMIX} on a sinkhole attack}
\label{arrival rates}
\end{figure}

We start by evaluating the ability of the \texttt{trustMIX} algorithm to route
messages to the sink in a network under a sinkhole attack (scenario 3)
and show that it succeeds in letting messages bypass the sinkholes. The left-hand side
of figure \ref{arrival rates} shows the instantaneous\footnote{Like in
section \ref{basic mix result}, we the curves are smoothened by taking the
average over $100$ time steps.} percentage of messages reaching the
sink for fixed radius $r=8$ and $d=4$, and when the
percentage of
attackers increases ($p=10$, $p=25$ and $p=50$). 
For $p=10$, we see
that \texttt{trustMIX} rapidly manages to safely deliver $80\%$ of the traffic
to the sink. The proportion of delivered traffic increases over time,
as the trust engine adjusts trust values for pairs of
nodes, and soon reaches about $90\%$. This should be compared to figure
\ref{capture rates}, where we see that \texttt{MIX} delivers only
about $10\%$ of the traffic safely, even for a small number of attackers
($p=10$). 
The comparison of the three curves on the left-hand side plot of
figure \ref{arrival rates} shows that the more attacker
nodes there are, the more messages get captured and the longer it takes for the
trust engine to adjust accurate trust values for pairs of
motes. However, even under harsh conditions ($p=50$, c.f. lower curve
of the left-hand side plot of figure \ref{arrival rates}), the
\texttt{trustMIX} algorithm is rapidly capable of delivering $50-60\%$ of the
traffic safely (again, compare this to figure \ref{capture rates},
where only about $10\%$ of the traffic reaches the sink for the \texttt{MIX}
algorithm even for $p=10$). On the
right-hand side of figure \ref{arrival rates}, we see that even for
$p=50$ the \texttt{trustMIX} algorithm manages to route safely about $80\%$ of
the traffic, but it needs more time to adjust correct trust values.
Summarizing, we see that \texttt{trustMIX} successfully attenuates the
effect of a sinkhole attack on the network by successfully delivering
an important fraction of the total traffic. 
\subsubsection{Impact of the Sinkhole Attack on Energy Consumption}
\begin{figure}[hbt]
\begin{center}
\includegraphics[angle=0,width=.3\textwidth]{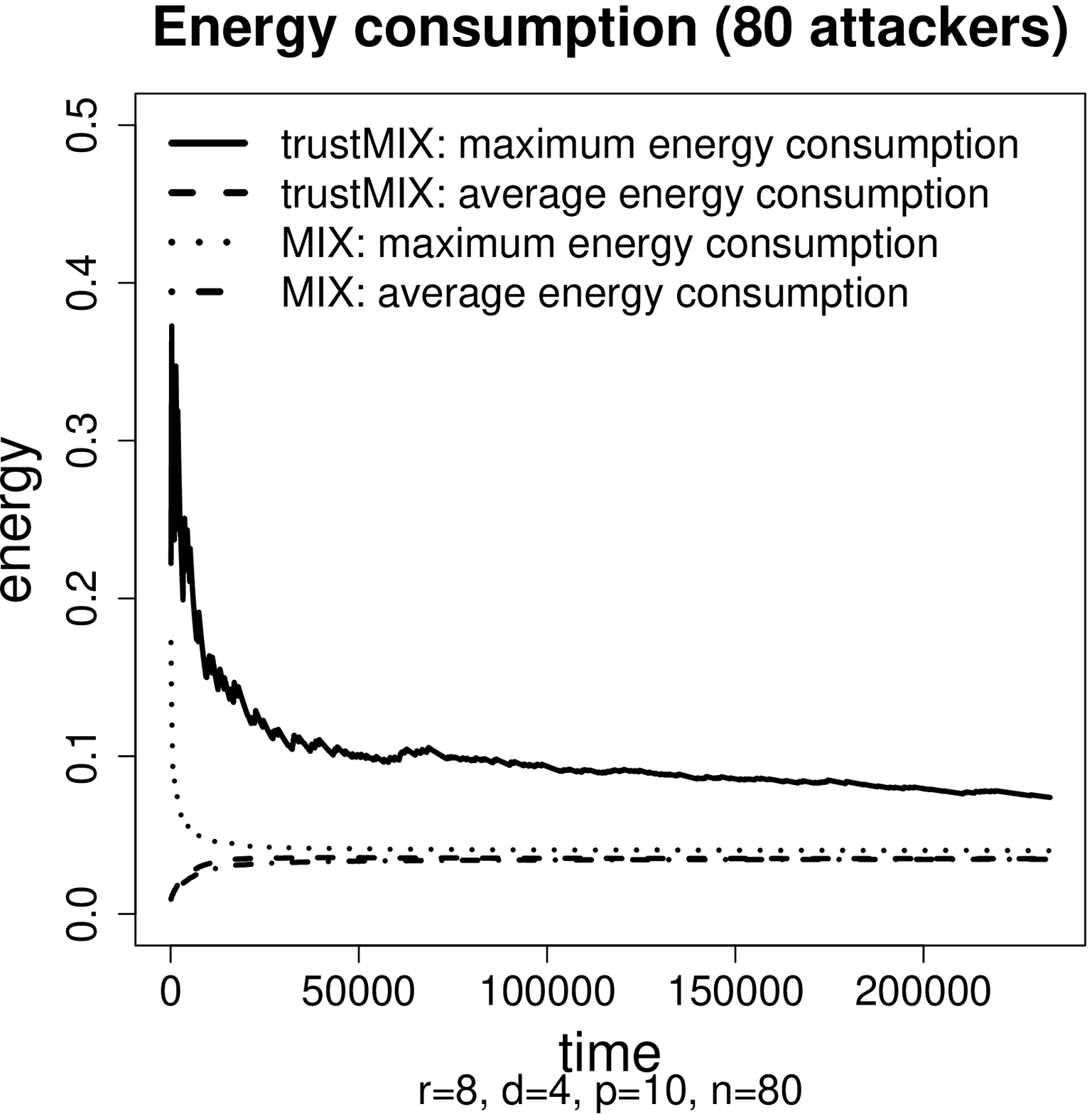}
\includegraphics[angle=0,width=.3\textwidth]{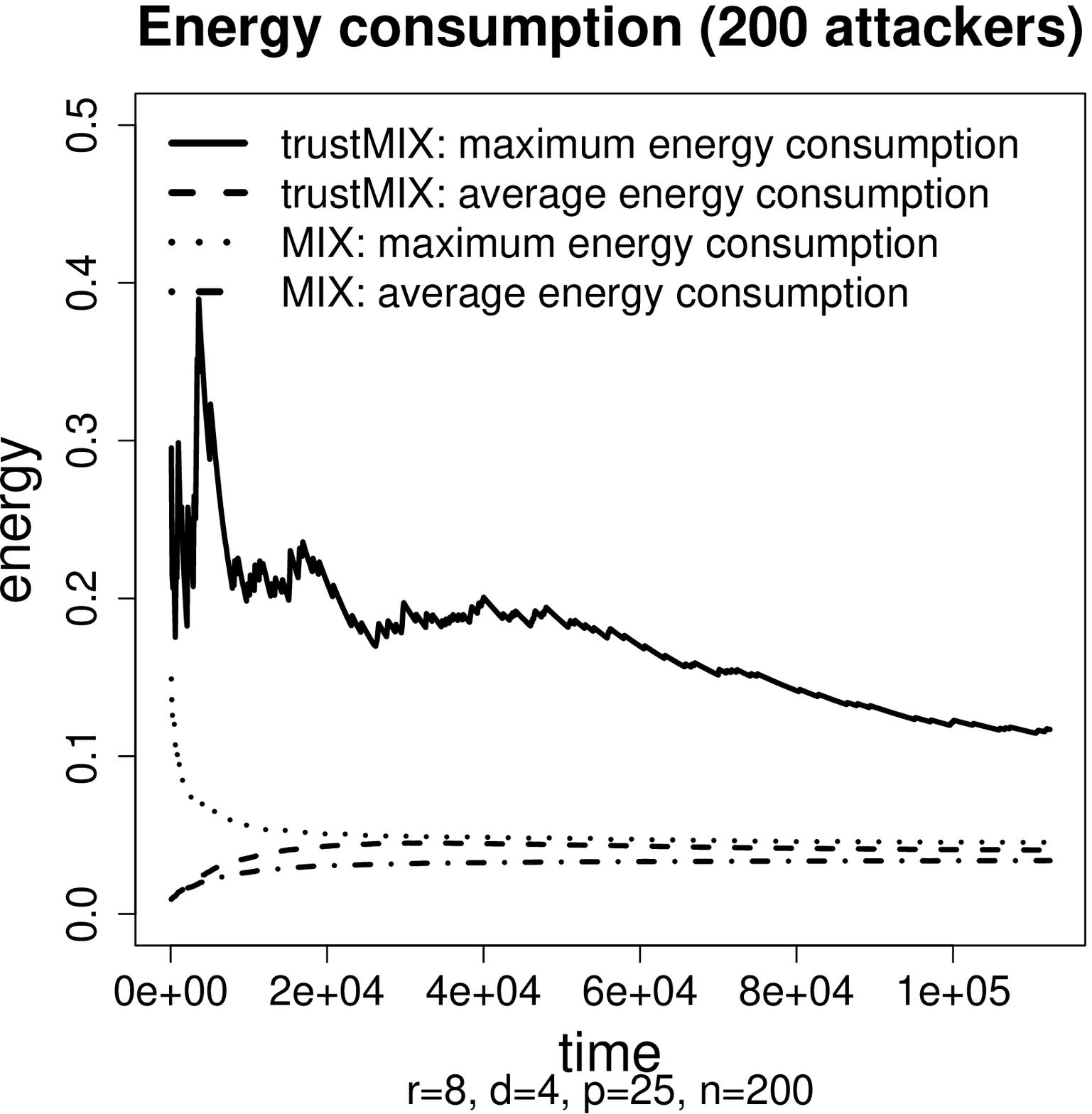}
\includegraphics[angle=0,width=.3\textwidth]{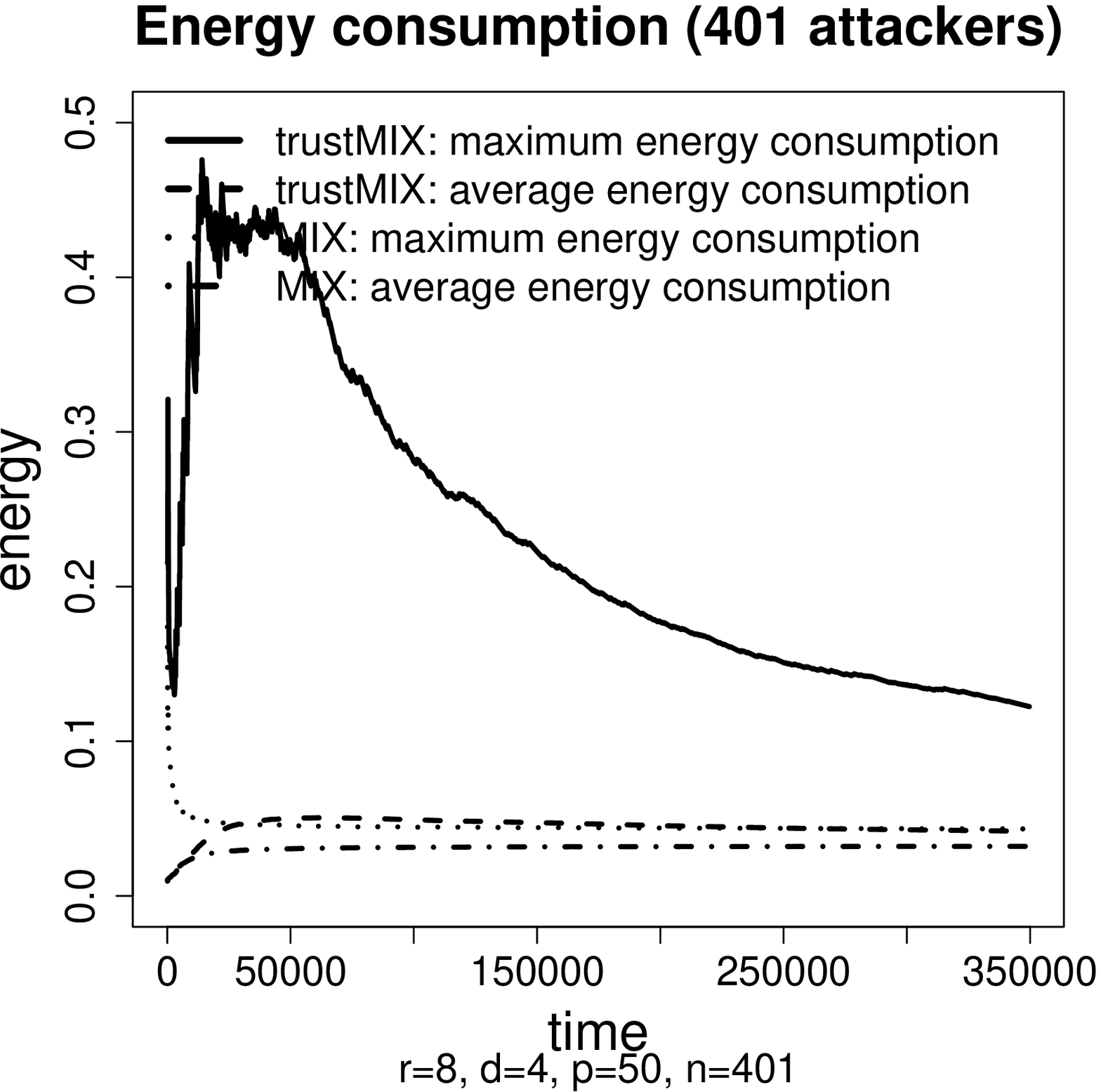}
\end{center}
\caption{Comparison of energy consumptions}
\label{nrg consumption}
\end{figure}

Next, we turn to evaluating the energy consumption of
\texttt{trustMIX} (scenario 3) in comparison to the energy consumption of
\texttt{MIX} in scenario 1.
\newtheorem{convention}{Convention}
\begin{convention}
In this section, it is implicit that \texttt{\textup{trustMIX}} is run with
\emph{scenario 3} and that \texttt{\textup{MIX}} is run with \emph{scenario 1}.
\end{convention}
We would like to point out that when a message is captured by a sinkhole, the network
stops consuming energy for routing it: this gives an unfair energy
advantage to \texttt{trustMIX}, which may have less
messages to route all the way to the sink than \texttt{MIX}.
Therefore, unless stated otherwise, we adopt the following convention
in this section:
\begin{convention}
\label{conv nrg is power}
When considering energy consumptions, it is implicit in this section
that we always consider energy consumptions per delivered message to
the sink and per time unit.
\end{convention} 
In figure \ref{nrg consumption}, we compare the average and the
maximum instantaneous\footnote{Again, taken as an average over $100$
  time steps} energy consumptions of \texttt{MIX} and \texttt{trustMIX}
 for different values of $p$. 
The average and maxima are taken over all sensor
nodes of the network.
As a preliminary remark, we observe that the \emph{maximum} and \emph{average} energy
consumptions of \texttt{MIX} are almost the same. This is consistent with
the finding \cite{JLPR06} that \texttt{MIX} evenly balances energy consumption among
sensors.

Next, if we look at the average energy consumption of \texttt{trustMIX}, we see that
it behaves very well for all tested values of $p$, since it is
almost the same as the average energy consumption of \texttt{MIX}. 
This means that \texttt{trustMIX} secures a network
under a sinkhole attack without spending more energy, on the average,
than \texttt{MIX} in a safe environment.

Looking at the maximum energy consumption of \texttt{trustMIX},
we see it is significantly greater than for \texttt{MIX}.
This is even more so when there are many attackers (central and right-hand side plots of figure \ref{nrg
  consumption}), and when the network has had little time to adjust
trust values (i.e. when $time$ is small). 
The first interpretation of this observation is that under a sinkhole attack,
the energy balancing feature of \texttt{MIX} is not well preserved by
\texttt{trustMIX} 
and some nodes spend more energy than others. This seems to be a weakness of
\texttt{trustMIX}, 
however a careful analysis shows that the key objective of
\texttt{MIX}, to increase
the lifespan of the network, is preserved by \texttt{trustMIX}
even under a sinkhole attack. In other words, we shall next explain why the maximum
energy consumption is not a good metric to measure the
lifetime of \texttt{trustMIX} under a sinkhole attack. On the
contrary, in
\cite{PLR06,JLPR06}, the maximum energy consumption is taken as a
measure of the lifetime of the network. The authors argue that the maximum
energy consumption is a good measure of lifetime for routing
algorithms where most of the traffic is sent hop-by-hop, including
\texttt{MIX}, but see \cite{PLR06} for details.

\begin{figure}[hbt]
\begin{center}
\includegraphics[angle=0,width=.3\textwidth]{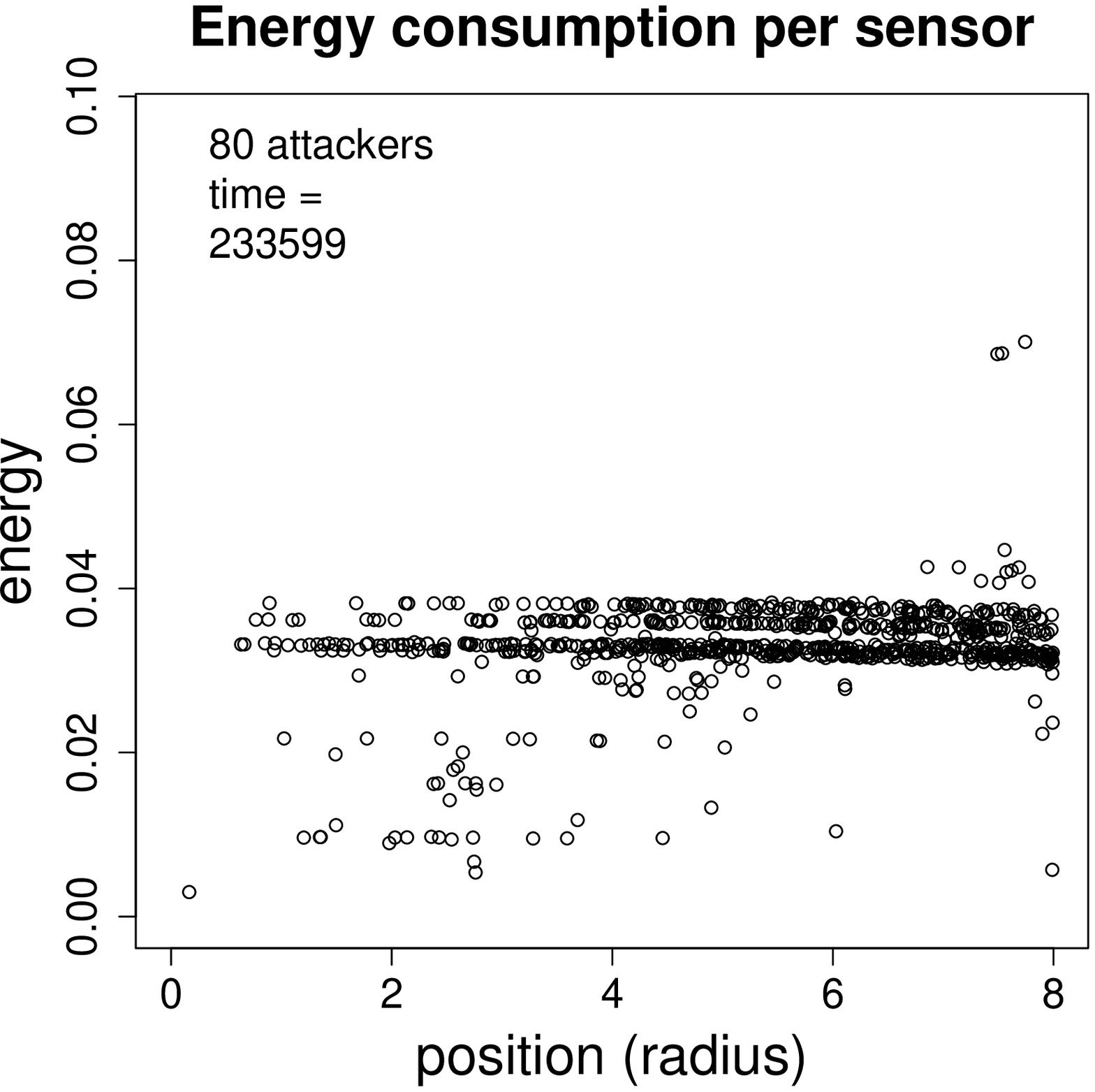}
\includegraphics[angle=0,width=.3\textwidth]{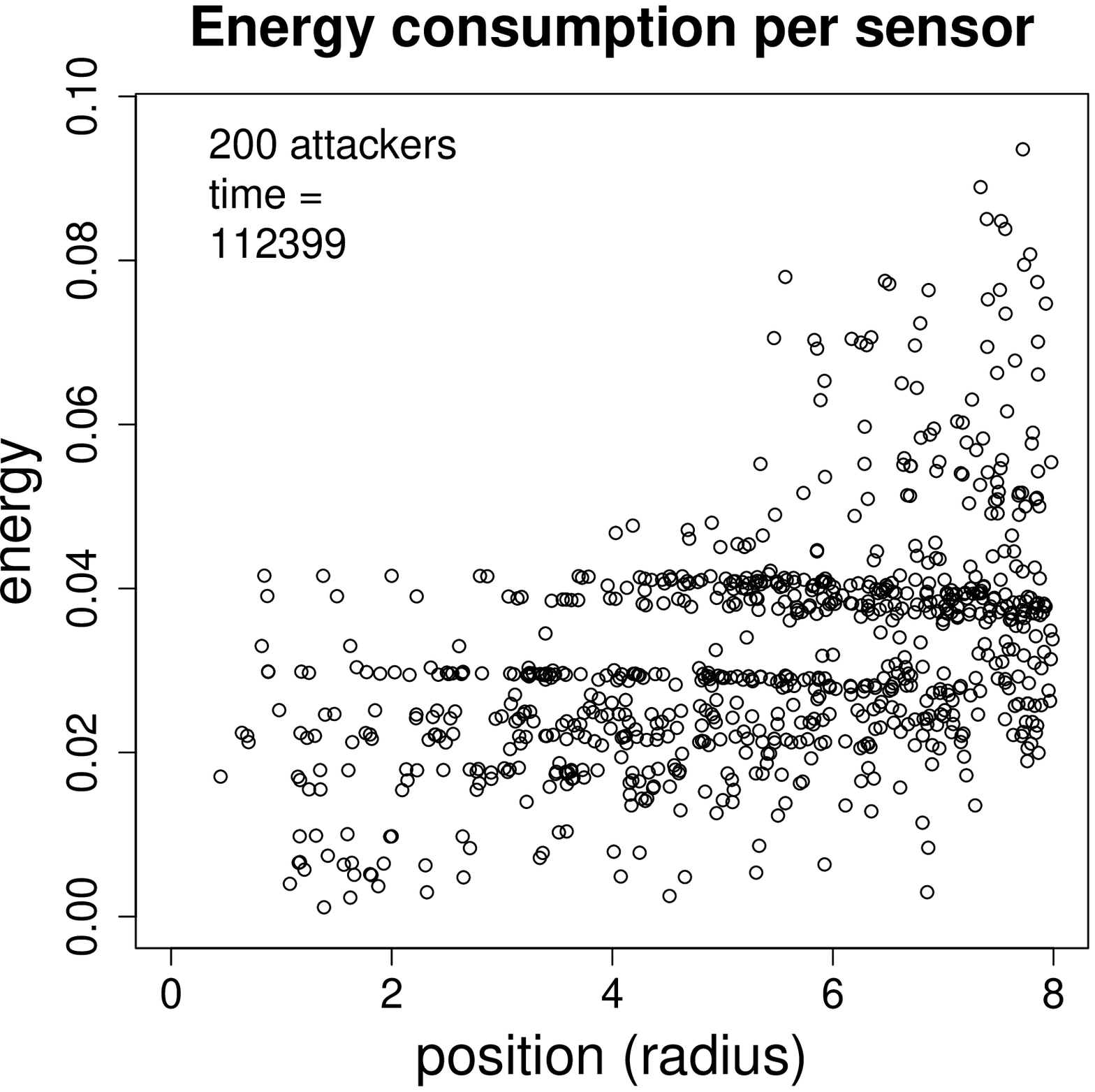}
\includegraphics[angle=0,width=.3\textwidth]{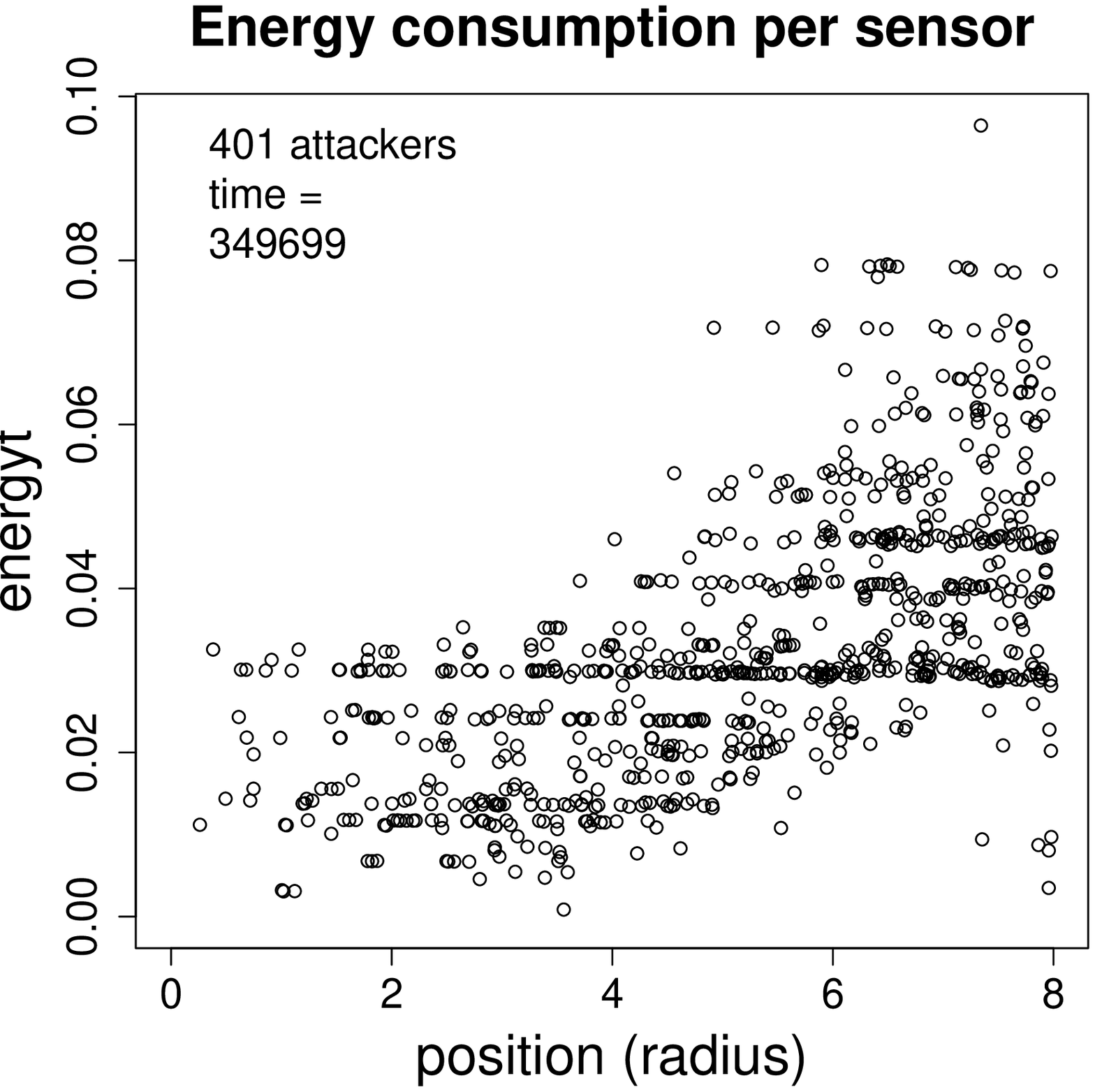}
\end{center}
\caption{Detailed observation of energy consumption for \texttt{trustMIX}}
\label{nrg consumption sensor-wise}
\end{figure}

In figure \ref{nrg consumption sensor-wise}, 
we look at the distribution of energy consumption in the network for
\texttt{trustMIX}, 
plotting the position of sensors (more precisely the distance
to the sink) against the total energy consumption (for each sensor)
divided by the total elapsed time, i.e. this is the average power,
for each sensor, in Joules per seconds (or Energy units per second,
c.f. section \ref{network model}). Please notice that 
in figure \ref{nrg consumption sensor-wise} we do
not follow convention \ref{conv nrg is power}  of dividing the energy by the number of
received messages.

As a first observation, we see that \emph{low energy consumption is preserved for
motes close to the sink}  for all values of $p$. 
Furthermore, when there are just a
few attackers ($p=10$) 
almost all nodes are energy balanced (left-hand side plot of
figure \ref{nrg consumption sensor-wise}). 

On the contrary, when the
network is under a sinkhole attack performed simultaneously by many
attackers ($p=25$ and $p=50$, center and right-hand side plots of
figure \ref{nrg consumption sensor-wise}), nodes \emph{away from the sink
deplete their energy reasonably but noticeably faster} than those close to the sink. 

However, unlike
the early energy depletion of nodes close to the sink that occurs for
hop-by-hop routing strategies (c.f. section \ref{Related work: energy saving}), the
energy depletion of motes away from the sink does not imply
disconnection of the sink from the
network and thus does not put the whole network down. 
Instead, when the network is
under attack by a high number of attacker motes, sensors away from the
sink will deplete energy slightly
faster, thus reducing the sensing range of the whole network but
\emph{without actually diminishing the network lifespan}, since as
seen on  figure \ref{nrg consumption}, the average energy consumption
of \texttt{trustMIX} is similar to the one of \texttt{MIX}.
\section{Conclusion}
\label{conclusion}
We have proposed \texttt{trustMIX}, a fully distributed and simple algorithm
for wireless sensor networks which extends and secures the \texttt{MIX}
algorithm from \cite{JLPR06} when submitted to massive sinkhole
attacks. 
Our three main results are to show 
(1) that the algorithm successfully lets most of the data avoid sinkholes
(2) preserving the main features of \texttt{MIX}: increased
lifespan of the network. We observe (3) that the mix TRUST algorithm
trades off
energy consumption from the motes situated on the outer boarder of the
network for increased security. In other words, if the network becomes subject to a
sinkhole attack, it successfully avoids the
sinkholes for most messages, but slowly becomes blind at its
peripheral, as nodes away from the sink deplete their energy
faster. This phenomenon seems to have less impact when the attacker
motes are not too numerous ($p=10$ in our simulations). 
We also see that time is a crucial factor, since the trust
mechanism needs time to detect the attacker motes, and the higher the
percentage of attacker nodes, the more time the trust engine needs to
secure the network. 
\section{Future Work}
\label{future work}
The \texttt{trustMIX} mechanism we propose shows that \texttt{MIX} can be efficiently
secured in the scenarios we have considered. Future work will analyse
parameters which have not been considered in this paper. Such
parameters include the deployment of attackers in a non-uniform manner
(for example, a region of the network could be attacked or attackers
could be deployed in strategic regions, e.g. close to the
sink). Also, we have only considered the ratio of delivered to
captured messages in the case of uniform events generation, but it
would be nice to investigate whether messages away from the sink are
more likely to get captured than those close to the sink. This will
eventually lead future work to propose enhanced versions of the
\texttt{trustMIX} algorithm we propose in this paper. For example, if messages away
from the sink are more likely to be captured, a multi-path strategy
could be applied to guarantee a lower percentage of captured
messages. The amount of message redundancy would be dynamically
computed taking into account parameters such as the distance to the
sink, the trust levels of neighbours and possibly the importance of
the message: if a message is labelled very important, it may be worth
spending more energy to ensure higher probability of delivery.
\bibliography{trustMix}		

\begin{thebibliography}{10}

\bibitem{mix}
Olivier Powell, Aubin Jarry, Pierre Leone, and Jose Rolim.
\newblock Gradient based routing in wireless sensor networks: a mixed strategy.
\newblock {\em arXiv.org automated e-print archives}, 2005.
\newblock Report CS-0511083.

\bibitem{JLPR06}
Aubin Jarry, Pierre Leone, Olivier Powell, and Jose Rolim.
\newblock An optimal data propagation algorithm for maximizing the lifespan of
  sensor networks.
\newblock In {\em IEEE International Conference of Distributed Computing in
  Sensor Systems (DCOSS)}, 2006.

\bibitem{PLR05}
Olivier Powell, Pierre Leone, and Jose Rolim.
\newblock Energy optimal data propagation in wireless sensor networks.
\newblock {\em arXiv.org automated e-print archives}, 2005.
\newblock Report CS-0508052. Journal version in Journal of Paralel and
  Distributed Computing.

\bibitem{PLR06}
Olivier Powell, Pierre Leone, and Jose Rolim.
\newblock Energy optimal data propagation in sensor networks.
\newblock {\em Journal of Parallel and Distributed Computation}, 67:302--317,
  2006.

\bibitem{sinkhole}
A~A Pirzada and C~S McDonald.
\newblock Circumventing sinkholes and wormholes in ad-hoc wireless networks.
\newblock 2005.

\bibitem{natureOfTrust}
D~Romano.
\newblock {\em The Nature of Trust: Conceptual and Operational Clarification}.
\newblock PhD thesis, Louisiana State University, 2003.

\bibitem{trustcomp}
Trustcomp online community.
\newblock http://www.trustcomp.com.

\bibitem{marsh94formalising}
S.~Marsh.
\newblock {\em Formalising Trust as a Computational Concept}.
\newblock PhD thesis, Department of Mathematics and Computer Science,
  University of Stirling, 1994.

\bibitem{despotovic}
Zoran Despotovic and Karl Aberer.
\newblock Trust and reputation management in p2p networks, July 6-9, 2004.

\bibitem{jmTrust}
JM~Seigneur.
\newblock {\em Trust, Security and Privacy in Global Computing}.
\newblock PhD thesis, Trinity College Dublin, 2005.

\bibitem{hu02wormhole}
Y.~Hu, A.~Perrig, and D.~Johnson.
\newblock Wormhole detection in wireless ad hoc networks, 2002.

\bibitem{sybil}
James Newsome, Elaine Shi, Dawn Song, and Adrian Perrig.
\newblock The sybil attack in sensor networks: analysis \& defenses.
\newblock In {\em IPSN '04: Proceedings of the third international symposium on
  Information processing in sensor networks}, pages 259--268, New York, NY,
  USA, 2004. ACM Press.

\bibitem{ebs}
Mohammed~A. Moharrum and Mohamed Eltoweissy.
\newblock A study of static versus dynamic keying schemes in sensor networks.
\newblock In {\em PE-WASUN '05: Proceedings of the 2nd ACM international
  workshop on Performance evaluation of wireless ad hoc, sensor, and ubiquitous
  networks}, pages 122--129, New York, NY, USA, 2005. ACM Press.

\bibitem{random}
H.~Chan, A.~Perrig, and D.~Song.
\newblock Random key predistribution schemes for sensor networks, 2003.

\bibitem{hwang04energymemorysecurity}
David~D. Hwang, Bo-Cheng~Charles Lai, and Ingrid Verbauwhede.
\newblock Energy-memory-security tradeoffs in distributed sensor networks.
\newblock In {\em ADHOC-NOW 2004}, volume 3158 of {\em LNCS}, 2004.

\bibitem{reputationGaneriwal}
Saurabh Ganeriwal and Mani~B. Srivastava.
\newblock Reputation-based framework for high integrity sensor networks.
\newblock In {\em SASN '04: Proceedings of the 2nd ACM workshop on Security of
  ad hoc and sensor networks}, pages 66--77, New York, NY, USA, 2004. ACM
  Press.

\bibitem{core}
Pietro Michiardi and Refik Molva.
\newblock Core: a collaborative reputation mechanism to enforce node
  cooperation in mobile ad hoc networks.
\newblock In {\em Proceedings of the IFIP TC6/TC11 Sixth Joint Working
  Conference on Communications and Multimedia Security}, pages 107--121,
  Deventer, The Netherlands, 2002. Kluwer, B.V.

\bibitem{confidant}
Sonja Buchegger and Jean-Yves~Le Boudec.
\newblock Performance analysis of the confidant protocol.
\newblock In {\em MobiHoc '02: Proceedings of the 3rd ACM international
  symposium on Mobile ad hoc networking \& computing}, pages 226--236, New
  York, NY, USA, 2002. ACM Press.

\bibitem{rfsn}
Saurabh Ganeriwal and Mani~B. Srivastava.
\newblock Reputation-based framework for high integrity sensor networks.
\newblock In {\em SASN '04: Proceedings of the 2nd ACM workshop on Security of
  ad hoc and sensor networks}, pages 66--77, New York, NY, USA, 2004. ACM
  Press.

\bibitem{johnson94routing}
D.~Johnson.
\newblock Routing in ad hoc networks of mobile hosts.
\newblock In {\em Workshop on Mobile Computing Systems and Applications}, Santa
  Cruz, CA, U.S., 1994.

\bibitem{RAS+00}
Jan~M. Rabaey, M.~Josie Ammer, Julio~L. da~Silva, Danny Patel, and Shad Roundy.
\newblock Picoradio supports ad hoc ultra-low power wireless networking.
\newblock {\em Computer}, 33(7):42--48, 2000.

\bibitem{WLLP01}
Brett Warneke, Matt Last, Brian Liebowitz, and Kristofer~S.J. Pfister.
\newblock Smart dust: Communicating with a cubic-millimeter.
\newblock {\em computer}, 34(1):44--51, 2001.

\bibitem{SL05}
Michael~J. Sailor and Jamie~R. Link.
\newblock "smart dust": Nanostructures devices in a grain of sand.
\newblock {\em Chemical Communication}, 11:1375--1383, 2005.

\bibitem{IGE00}
C.~Intanagowiwat, R.~Govindan, and D.~Estrin.
\newblock Directed diffusion: A scalable and robust communication paradigm for
  sensor networks.
\newblock In {\em 6th International Conference on Mobile Computing (MOBICOM)}.
  ACM/IEEE, 2000.

\bibitem{CNS02}
I.~Chatzigiannakis, S.~Nikoletseas, and P.~Spirakis.
\newblock Smart dust protocols for local detection and propagation.
\newblock In {\em 2nd Workshop on Principles of Mobile Computing (POMC)}, pages
  9--16. ACM, ACM Press, 2002.

\bibitem{HCB00}
W.R. Heinzelman, A.~Chandrakasan, and H.~Balakrishnan.
\newblock Energy efficient communication protocol for wireless microsensor
  networks.
\newblock In {\em 33th Hawaii International Conference on System Sciences
  (HICSS)}, 2000.

\bibitem{STGS02}
C.~Schurgers, V.~Tsiatsis, S.~Ganeriwal, and M.~Srivastava.
\newblock Topology management for sensor networks: Exploiting latency and
  density.
\newblock In {\em 8th International Conference on Mobile Computing (MOBICOM)}.
  ACM/IEEE, 2002.

\bibitem{CT00}
J.~Chang and L.~Tassiulas.
\newblock Energy conserving routing in wireless ad hoc networks.
\newblock {\em IEEE INFOCOM}, 1:22--31, 2000.

\bibitem{LSS05}
L.~Lin, N.B. Shroff, and R.~Srikant.
\newblock {Asymptotically optimal power-aware routing for multihop wireless
  networks with renewable energy sources}.
\newblock {\em Proceedings of INFOCOM'05}, 2005.

\bibitem{HGWC02}
X.~Hong, M.~Gerla, H.~Wang, and L.~Clare.
\newblock {Load balanced, energy-aware communications for Mars sensor
  networks}.
\newblock {\em Aerospace Conference Proceedings, 2002. IEEE}, 3:3--1109, 2002.

\bibitem{BCN05}
A.~Boukerche, I.~Chatzigiannakis, and S.~Nikoletseas.
\newblock {Power-Efficient Data Propagation Protocols for Wireless Sensor
  Networks}.
\newblock {\em SIMULATION}, 81(6):399--411, 2005.

\bibitem{L06}
J.~Luo.
\newblock {\em Mobility in Wireless networks: friend or Foe - network design
  and Control in the Age of Mobile Computing}.
\newblock PhD thesis, School of Computer and Communications Science, EPFL,
  Switzerland, 2006.

\bibitem{LH05}
J.~Luo and J.-P. Hubaux.
\newblock Joint mobility and routing for lifetime elongation in wireless sensor
  networks.
\newblock In {\em 24th INFOCOM}, 2005.

\bibitem{LH06}
J.~Luo and J.-P. Hubaux.
\newblock Mobility to improve the lifetime of wireless sensor networks: A
  theoretical framework.
\newblock In {\em Workshops of the Second International Conference on
  Distributed Computing in Sensor Systems (DCOSS)}, 2006.

\bibitem{SD05}
M.L. Sichitiu and R.~Dutta.
\newblock {Benefits of Multiple Battery Levels for the Lifetime of Large
  Wireless Sensor Networks}.
\newblock In {\em NETWORKING 2005: 4th International IFIP-TC6 Networking
  Conference}, Lecture Notes in Computer Science, pages 1440--1444. Springer
  Berlin/Heidelberg, May 2005.

\bibitem{OS06}
S.~Olariu and I.~Stojmenovic.
\newblock {Design Guidelines for Maximizing Lifetime and Avoiding Energy Holes
  in Sensor Networks with Uniform Distribution and Uniform Reporting}.
\newblock In {\em 25th Conference on Computer Communications (INFOCOM)}. IEEE
  Communications Society, IEEE Computer Society Press, April 2006.

\bibitem{B05}
Azzedine Boukerche.
\newblock {\em Handbook of Algorithms for Wireless Networking and Mobile
  Computing}.
\newblock Chapman \& Hall/CRC, 2005.

\bibitem{BN04}
A.~Boukerche and S.~Nikoletseas.
\newblock {\em Wireless Communications Systems and Networks}, chapter Protocols
  for Data Propagation in Wireless Sensor Networks: A Survey, pages 23--51.
\newblock Kluwer Academic Publishers, 2004.

\bibitem{AK05}
Jamal~N. Al-Karaki and Ahmed~E. Kamal.
\newblock A taxonomy of routing techniques in wireless sensor networks.
\newblock In Mohammad Ilyas and Imad Mahgoub, editors, {\em Handbook of Sensor
  Networks: Compact Wireless and Wired Sensing Systems}, pages 6.1--6.24. CRC
  Press, 2005.

\bibitem{AY05}
Kemal Akkaya and Mohamed Younis.
\newblock A survey on routing protocols for wireless sensor networks.
\newblock {\em Ad Hoc Network Journal}, 3/3:325--349, 2005.

\bibitem{SP03}
M.~Singh and V.~Prasanna.
\newblock Energy-optimal and energy-balanced sorting in a single-hop wireless
  sensor network.
\newblock In {\em First International Conference on Pervasive Computing and
  Communications (PERCOM)}. IEEE, 2003.

\bibitem{YP03}
Y.~Yu and V.K. Prasanna.
\newblock Energy-balanced task allocation for collaborative processing in
  networked embedded system.
\newblock In {\em Conference on Language, Compilers, and Tools for Embedded
  Systems (LCTES)}, June 2003.

\bibitem{GLW03}
W.~Guo, Z.~Liu, and Guangbin Wu.
\newblock An energy-balanced transmission scheme for sensor networks.
\newblock In {\em Poster Session of the First International Conference on
  Embedded Networked Sensor Systems (SenSys)}. ACM, IEEE Computer Society
  Press, November 2003.

\bibitem{LXG05}
Z.~Liu, D.~Xiu, and Weihua Guo.
\newblock An energy-balanced model for data transmission in sensor networks.
\newblock In {\em 62nd Semiannual Vehicular Technology Conference}, September
  2005.

\bibitem{ENR04}
C.~Efthymiou, S.~Nikoletseas, and J.~Rolim.
\newblock Energy balanced data propagation in wireless sensor networks.
\newblock In {\em Best papers of the 4th International Workshop on Algorithms
  for Wireless, Mobile, Ad Hoc and Sensor Networks}, 2004.

\bibitem{ENR06}
C.~Efthymiou, S.~Nikoletseas, and J.~Rolim.
\newblock Energy balanced data propagation in wireless sensor networks.
\newblock {\em Wireless Networks (WINET) Journal}, 2006.

\bibitem{LNR05}
P.~Leone, S.~Nikoletseas, and J.~Rolim.
\newblock An adaptive blind algorithm for energy balanced data propagation in
  wireless sensor networks.
\newblock In {\em The First International Conference on Distributed Computing
  in Sensor Systems (DCOSS)}, number 3560 in Lecture Notes in Computer Science.
  Springer Verlag, June/July 2005.

\bibitem{PN07}
Olivier Powell and Sotiris Nikoletseas.
\newblock Simple and efficient geographic routing around obstacles for wireless
  sensor networks.
\newblock In {\em Workshop on Experimental Algorithms (WEA)}, pages 161--174.
  Springer-Verlag, 2007.

\bibitem{PN07b}
Olivier Powell and Sotiris Nikolesteas.
\newblock Geographic routing around obstacles in wireless sensor networks.
\newblock Technical Report cs.DC/0703094, Computer Science ArXiv of
  Distributed, Parallel and Cluster Computing, Mars 2007.

\bibitem{LMPR06}
Pierre Leone, Luminita Moraru, Olivier Powell, and Jose Rolim.
\newblock A localization algorithm for wireless ad-hoc sensor networks with
  traffic overhead minimization by emission inhibition.
\newblock In {\em Proceedings of ALGOSENSOR'06}, 2006.

\end{thebibliography}
\bibliographystyle{unsrt}

\end{document}